# Galactic optical cloaking of visible baryonic matter

Igor I. Smolyaninov

*Department of Electrical and Computer Engineering, University of Maryland, College Park, MD 20742, USA*

**Three-dimensional gravitational cloaking is known to require exotic matter and energy sources, which makes it arguably physically unrealizable. On the other hand, typical astronomical observations are performed using one-dimensional paraxial line of sight geometries. We demonstrate that unidirectional line of sight gravitational cloaking does not require exotic matter, and it may occur in multiple natural astronomical scenarios which involve gravitational lensing. In particular, recently discovered double gravitational lens SDSSJ0946+1006 together with the Milky Way appear to form a natural paraxial cloak. A natural question to ask, then, is how much matter in the universe may be hidden from view by such natural gravitational cloaks. It is estimated that the total volume hidden from an observer by gravitational cloaking may reach about 1% of the total volume of the visible universe.**

Recent developments in metamaterials and transformation optics, which partially rely on general relativity [1], have sparked considerable interest in invisibility cloaking. Several theoretical cloaking schemes had been proposed [2,3], which were quickly followed by experimental demonstrations in various portions of electromagnetic spectrum [4-6]. This body of work, which was done using various engineered effective metrics in "virtual" electromagnetic space-time, has inspired an effort to determine if gravitational cloaking can be achieved as a result of the gravitational curvature of physical space-time [7]. If possible, such gravitational cloaking would potentially shed a new light on the issue of dark matter in the universe. Unfortunately, it appears that full three-dimensional gravitational cloaking requires exotic matter and energy sources [7], which makes it physically unrealizable.

While this conclusion will probably hold true in the 3D case, recent advances in cloaking technology led to the development of several simplified cloaking schemes. One of these schemes uses arrangements of conventional glass lenses (see Fig.1) assembled into one-dimensional light of sight invisibility cloaks [8]. Since typical astronomical observations are also performed using various one-dimensional line of sight arrangements, the issue of feasibility of gravitational cloaking needs to be re-examined. It appears that such unidirectional line of sight cloaking does not require any exotic matter, and it may occur in multiple natural astronomical scenarios which involve gravitational lensing. The problem of missing (dark) matter in the universe is one of the most important problems in contemporary physics. The described novel cloaking mechanism is a large effect, which may account for considerable portion of this missing mass. It is estimated below that the total volume hidden from an observer by gravitational cloaking may reach at least 1% of the total volume of the visible universe. Moreover, this estimate may potentially be revised upwards depending on the estimated total number of galaxies in the universe.

First, let us consider the line of sight cloaking geometries shown in Fig.1(a,b). Compared to [8], these geometries are simplified due to the absence of chromatic aberrations in vacuum. These cloaking devices are based on ray optics and operate in the paraxial (small angle limit) approximation. The four-lens cloak is made of two pairs of positive lenses arranged in such a way that the focal spots of the neighbouring lenses in each pair coincide with each other, while the three-lens cloak requires a negative lens positioned in between two positive lenses. In the paraxial limit these cloaks do not change the ray angles exiting the device. The image of an object behind the cloak, as seen by an observer, is identical to the object itself. Nevertheless, a considerable volume inside the cloaks is completely hidden from an observer positioned along the optical axis. It is important to note that cloaking behaviour in such lens assemblies is quite stable with respect to broad variations of parameters of the individual lenses. Indeed the Choi-Howell configuration allows considerable variations in lens parameters as long as the focal points of the neighbouring lenses coincide. For example, while the original configuration presented in [8] is symmetric with respect to the central plane (see Fig.1(a)), the two central lenses and the two side lenses do not need to be similar. In fact, all four lenses may be different as long as the focal points of the neighbouring lenses coincide. Such flexibility of parameters is important if cloaking behaviour is to be observed in random lens arrangements, such as various arrangements of galaxy-scale gravitational lenses. Unfortunately, as demonstrated in [8], such ideal ray optics cloaks are possible only if three or more lenses are used. However, as illustrated in Fig.1(c) it is still possible to hide objects in between two lenses in the paraxial ray optics limit. Such a two-lens hiding arrangement would invert ray positions, while keeping the size, shape, and colour of the object behind the lenses. Since the background distribution of matter is often not known, inversion of the background image may not be easily detectable. As illustrated in Fig.1(d), this line of sight paraxial cloaking or "hiding" may also be quite efficient. According to [8], the measured field of view of their cloak was about 3 degrees, which is very large by astronomical standards.

Let us now determine what kind of astronomical objects may be involved in natural gravitational cloaking. The local value of the effective index of refraction $n$ of a gravitational lens is determined by gravitational potential $\Phi$ [9]:

$$n(r) \approx 1 - \frac{2\Phi(r)}{c^2} \tag{1}$$

Therefore, real gravitational lenses may exhibit quite complicated spatial distributions of $n(r)$. However, these distributions may be simplified in several cases. For a point mass $M$ the gravitational deflection angle of light passing a massive body is [9]:

$$\theta = \frac{4\gamma M}{rc^2} = \frac{2r_g}{r}, \qquad (2)$$

where $r_g$ is its gravitational radius. In typical astronomical situations the physical size of the gravitational lens is much smaller than the distances between an observer, a lens and a source. This justifies the usage of the thin screen approximation in which the lens is approximated by a planar distribution of matter in the "lens plane". If the mass distribution inside such a planar lens is axially symmetric, the gravitational deflection angle may be written as

$$\theta(r) = \frac{4\gamma M(r)}{rc^2}, \qquad (3)$$

where $M(r)$ is the mass enclosed by the circle of radius $r$. Based on Eq.(3), the distance $F$ to the gravitational focus may be found as

$$F(r) = \frac{r^2}{\left(\frac{4\gamma M(r)}{c^2}\right)} \qquad (4)$$

[10]. Determination of $M(r)$ in a given galaxy is complicated by an unknown amount of dark matter. If the surface mass density within an axially symmetric planar galactic lens is approximately constant, the focal distance $F$ does not depend on $r$ within the lens, which would lead to ideal focusing. However, the surface mass density typically decreases as a function of radius, so that ray focusing is not ideal, as shown in Fig.2(a). Nevertheless, since in spiral galaxies the rotation curves are almost flat, the effective refractive index distribution, which may be obtained directly from the measured galaxy rotation curve as

$$v^2 = r\frac{d\Phi}{dr} \tag{5}$$

appears to be logarithmic and reasonably close to that of a good positive lens, which is evidenced by multiple observations of very distant and otherwise unobservable galaxies aided by galaxy-scale gravitational lenses [11]. An extensive catalogue of various galaxy mass distribution models used in evaluation of their gravitational lensing properties is available in [12]. One of the most widely used models is the singular isothermal sphere profile illustrated in Fig.2(b). The spherically isotropic density profile in this model is

$$\rho(r) = \frac{\sigma_v^2}{2\pi\gamma r^2}, \tag{6}$$

where $\sigma_v$ is the velocity dispersion. Since the total mass calculated by integrating this function does not converge, it is cut off at some effective radius $R_{eff}$. The effective refractive index of this mass distribution calculated using Eq.(1) is constant:

$$n = 1 + \frac{4\sigma_v^2}{c^2}, \tag{7}$$

and its focal length may be determined using the well-known expression for a ball lens:

$$F = \frac{nR_{eff}}{2(n-1)} = \frac{R_{eff}c^2}{8\sigma_v^2} \tag{8}$$

As illustrated in Fig.2(b), in the paraxial approximation a ball lens "hides" a considerable volume of space around its focal spot. This hiding is never perfect due to the absence of "sharp edges" in real life gravitational lenses and the clearly visible distortion of the lens background. However, the visibility of objects in the hidden volume is strongly suppressed. We must therefore conclude that while such gravitational lenses as the galaxy cluster MACSJ0717.5+3745 shown in Fig.2(c) enable imaging of distant galaxies, they may also suppress visibility of considerable volumes of space around their focal spots. In the following we will demonstrate that visibility of these volumes may be further suppressed by the paraxial cloaking effects. We should also point out that most galaxy-scale gravitational lenses are weak, which means that in the first order approximation vast majority of these lenses may be described by their average effective parameters, such as the effective focal length *F*. The second order corrections due to effective refractive index non-homogeneities will be considerably weaker in most cases. Such an approach works successfully in other weak lensing situations. For example, it works well in theoretical descriptions of light propagation through turbulent atmosphere [13]. The turbulent atmosphere in these theories is successfully represented by a large number of weak (n~1) random lenses regardless of the exact details of mass distribution around atmospheric non-homogeneities.

The brief consideration above illustrates the well-known fact that gravitational lenses are abundant in the universe. Now we will show that paraxial gravitational cloaking based on chance arrangements of these lenses may indeed occur in multiple astronomical scenarios. Let us start with seemingly the least plausible scenario of a four-lens cloak shown in Fig.1(a). Such a cloak is supposed to be formed by four positive gravitational lenses which are well aligned with respect to each other along the same optical axis. Despite seeming improbability of such a chance arrangement, cosmic objects like this have been already observed in the experiment. Let us consider, for example, the recent double Einstein ring observation around the gravitational lens SDSSJ0946+1006 (see Fig.3(a)) reported in [14]. According to the analysis conducted in [14], this ring structure arises from the light from three galaxies located at distances of 3, 6, and 11 billion light years. As illustrated in Fig.3(b), together with the Milky Way galaxy these three galaxies form a four-lens arrangement aligned with each other along the same optical axis. Would it operate as a paraxial gravitational invisibility cloak for a distant observer located along the same optical axis? In order to answer this question we need to evaluate focal distances of the lenses in this four-lens arrangement.

Let us start with the Milky Way galaxy. Its gravitational lensing properties (as perceived by a distant observer) have been considered in [15]. In this work the total lensing potential $\Phi$ of the Milky Way was decomposed into separate contributions from its bulge (Eq.(2)), disc (Eq.(4)) and halo (Eqs.(6-8)), and its anisotropic lensing properties have been considered assuming different orientations of the Milky Way disk with respect to various hypothetic observes. While the bulge behaves like a compact mass with an effective radius of 0.7 kpc, for the calculated effective optical radii $R_{eff}$ of the disc and the halo the values of 6.5 kpc and 6.0 kpc have been reported, respectively. Since we only want to perform a basic evaluation of the Milky Way suitability for use in a four-lens cloaking arrangement, we will use the simplified Eq.(4) and determine the location of the Milky Way focal spot as

$$F = \frac{R_{eff}^2}{2r_g} \quad (9)$$

Based on [15], we may consider the Milky Way as a planar gravitational lens with an effective radius $R_{eff} \sim 6.0$ kpc$=2\times10^4$ ly. As far as its gravitational radius $r_g$ is concerned, there is an uncertainty in its value due to the yet unknown amount of dark matter within the Milky Way. A recent measurement based on the radial velocity of halo stars produced an estimate of $7\times10^{11}$ solar masses for the total mass of the Milky Way [16], which leads to $r_g=0.22$ ly. As a result, we obtain the focal length $F_{MW}=0.91$bly (billions of light years) for the Milky Way galaxy.

Now let us turn our attention to the gravitational lens SDSSJ0946+1006 which was analysed in detail in [14]. For the first galaxy in this lensing arrangement, which is located 3 bly from the Mlky Way, the effective lens radius $R_{eff}$ and the total mass have been estimated as 7.29 kpc $=2.4$x$104$ ly and $4.9$x$10^{11}$ solar masses, respectively [14]. Based on Eq.(9), the focal spot of SDSSJ0946+1006 is positioned at $F=1.9$ bly from this galaxy. Thus, according to these estimates it appears that the focal spots of the Milky

Way and the first galaxy in the SDSSJ0946+1006 system almost coincide with each other. This is illustrated in Fig.3(b). Therefore, based on Fig.1(c) and the arguments above, a considerable volume of space in between the Milky Way and the first galaxy in the SDSSJ0946+1006 system must be hidden from view of a distant observer located along the same optical axis. This finding is very important since it represents a demonstrable example of a gravitational lens-based hiding device.

While we know that the second and the third galaxies in the SDSSJ0946+1006 system (seen as two rings in Fig.3(a)) are located along the same optical axis as the Milky Way and the first galaxy [14], the resulting four-lens arrangement may or may not work as an efficient four-lens paraxial cloak, depending on whether the focal spots of the third and the fourth galaxies in this arrangement coincide with each other. The mass of the second galaxy was estimated in [14] as $1.5 \times 10^{11}$ solar masses based on its absolute magnitude and the Tully-Fisher relation [17], so that its gravitational radius appears to be $r_g$=0.047 ly. However, virtually nothing is known about its mass distribution, so its focal distance remains uncertain. Even less is known at present about the last galaxy in this arrangement. However, if we accept the Milky Way and the first galaxy in the SDSSJ0946+1006 system as typical examples, we may assume that the focal spot of a typical galaxy should be located several billion light years away. Therefore, the possibility that the Milky Way and the three galaxies in the SDSSJ0946+1006 system form a four-lens paraxial cloaking device cannot be excluded at present. Regardless of the ultimate resolution of this issue, from the consideration above it appears quite certain that the Milky Way and the first galaxy in the SDSSJ0946+1006 system should hide a large volume of space from a hypothetical distant observer located on the same optical axis (see Fig.3(b).

Let us also briefly consider the three-lens cloaking configuration shown in Fig.1(b). While cloaking is possible in such a configuration, the necessity to have a negative lens in between two positive lenses makes it quite improbable. Compared to positive lensing, negative gravitational lensing is believed to be a rare phenomenon, since it requires an inverse density profile, which goes against gravitational clumping of matter. It was hypothesized to occur in between clumps of matter, so that the matter density between the light rays is less than the overall background density [18]. An inverse density profile may be also produced in a supernova remnant and in such exotic situations as near the mouth of a wormhole [19]. While these situations cannot be completely discarded, let us concentrate on the much more probable hiding geometries based on either one or two positive gravitational lenses.

Since in most cases we do not exactly know what a given cosmic background looks like (owing to the fact that we only observe line-of-site views of most galaxies), there is no need to achieve complete "cloaking" in the usual sense of the word. As illustrated in Fig.2(b), even a single galaxy-scale gravitational lens may hide a considerable volume. The volume of space hidden from observation in the paraxial limit by "stand alone" galaxy-scale gravitational lenses (see Fig.2(b)) may be estimated based on the number of galaxies in the visible universe and on the volume hidden by a typical galaxy. The recent estimate in [20] reports the total number of galaxies at cosmological redshifts $z<8$ to be about $2 \times 10^{12}$. Let us assume that the Milky Way and the SDSSJ0946+1006 represent typical examples which may be used to evaluate a typical hidden volume behind a galaxy. Based on Fig.2(b), we may estimate the hidden volume as

$$V_{hidden} = \frac{4\pi}{3} R_{eff}^2 F \qquad (10)$$

If $R_{eff}=2.2 \times 10^4$ ly and $F=1.5 \times 10^9$ ly are assumed as typical values, the total volume hidden from an observer by $2 \times 10^{12}$ galaxies is about $5 \times 10^{30}$ ly$^3$, which is more than 1% of the $4 \times 10^{32}$ ly$^3$ volume of the visible universe. Since this estimate is based on quite a few simplifications, while the hidden volume appears to be quite large with respect to the total volume of the universe, the described gravitational cloaking effect should be important in the context of mass-energy balance in the universe. This estimate may potentially be revised upwards depending on the total number of galaxies in the universe.

As far as more perfect two-lens cloaking (see Fig.1(c)) is concerned, the number of such systems may be evaluated based on the number of strong gravitational lenses. As a result of a number of dedicated search efforts, hundreds of galaxy-scale gravitational lens systems are known at present. As estimated in [21], there should be ~20 strong gravitational lenses per square degree at space-based depth and resolution, which corresponds to at least $10^6$ such systems in the visible universe. However, the actual number of such two-lens cloaks may be smaller since the focal spots of the two galaxies in the two-lens arrangement need to match with each other.

We can also briefly evaluate if cloaking due to gravitational microlensing may occur on the sub-galactic scale. The gravitational focus of the Sun is located at a distance of about 0.01 ly [10], which is considerably smaller than typical distances between stars in the Sun neighbourhood. However, at the cores of globular star clusters the typical distances between stars may be as small as $\sim 10^{-2}$-$10^{-3}$ ly [22], which makes various cloaking arrangements from Fig.1 plausible. Since globular star clusters play a very important role in cosmic distance ladder, evaluation of potential cloaking effects on various visual characteristics of the globular clusters would be very important.

In conclusion, we have demonstrated that unidirectional line of sight gravitational cloaking does not require exotic matter, and it may occur in multiple natural astronomical scenarios which involve gravitational lensing. In particular, recently discovered double gravitational lens SDSSJ0946+1006 together with the Milky Way appear to form a natural paraxial cloak. It is estimated that the total volume hidden from an observer by gravitational cloaking may reach about 1% of the total volume of the visible universe, which is important in the context of mass-energy balance in the universe.

**Figure Captions**

**Figure 1.** (a) Schematic geometry of a four-lens cloak made of four positive lenses. Ray propagation through the cloak is shown for three rays (red, green and blue). The cloaked volume is shaded in grey. (b) Schematic geometry of a three-lens cloak made of a negative lens in between two positive lenses. (c) While a two-lens paraxial cloak is impossible to make, an arrangement of two positive lenses is also capable of hiding considerable volume (shaded in grey). This arrangement inverts ray positions, while keeping the size, shape, and colour of the object behind the lenses. (d) Example of the paraxial cloak in action from ref.[8].

**Figure 2**. (a) Comparison of gravitational lensing by a galaxy having constant surface mass density (left) and a galaxy in which surface mass density decreases as a function of radius (right). (b) One of the most widely used galaxy mass models utilized in evaluation of their gravitational lensing properties is the singular isothermal sphere (SIS) profile. Note that in the paraxial approximation such a gravitational lens hides a considerable volume of space around its focal spot. (c) Image of the galaxy cluster MACSJ0717.5+3745. It is one of the most massive galaxy clusters known, and it is also the largest known gravitational lens. While such a gravitational lens enables imaging of distant galaxies, based on schematic diagram shown in (b), it may also hide considerable volume of space around its focal spot. Image credit: NASA/ESA

**Figure 3**. (a) Using the Hubble Space Telescope, a double Einstein ring SDSSJ0946+1006 has been found in observations reported in [14]. This ring structure arises from the light from three galaxies at distances of 3, 6, and 11 billion light years. Image credit HST/NASA/ESA. (b) Together with our own Milky Way galaxy, for a distant observer located along the same optical axis these three galaxies form a four-lens

arrangement aligned with each other along the same optical axis. Since their focal spots (shown in the respective colour) coincide with each other, the Milky Way and the first lens in the SDSSJ0946+1006 arrangement form a paraxial hiding device similar to one shown in Fig.1(c).

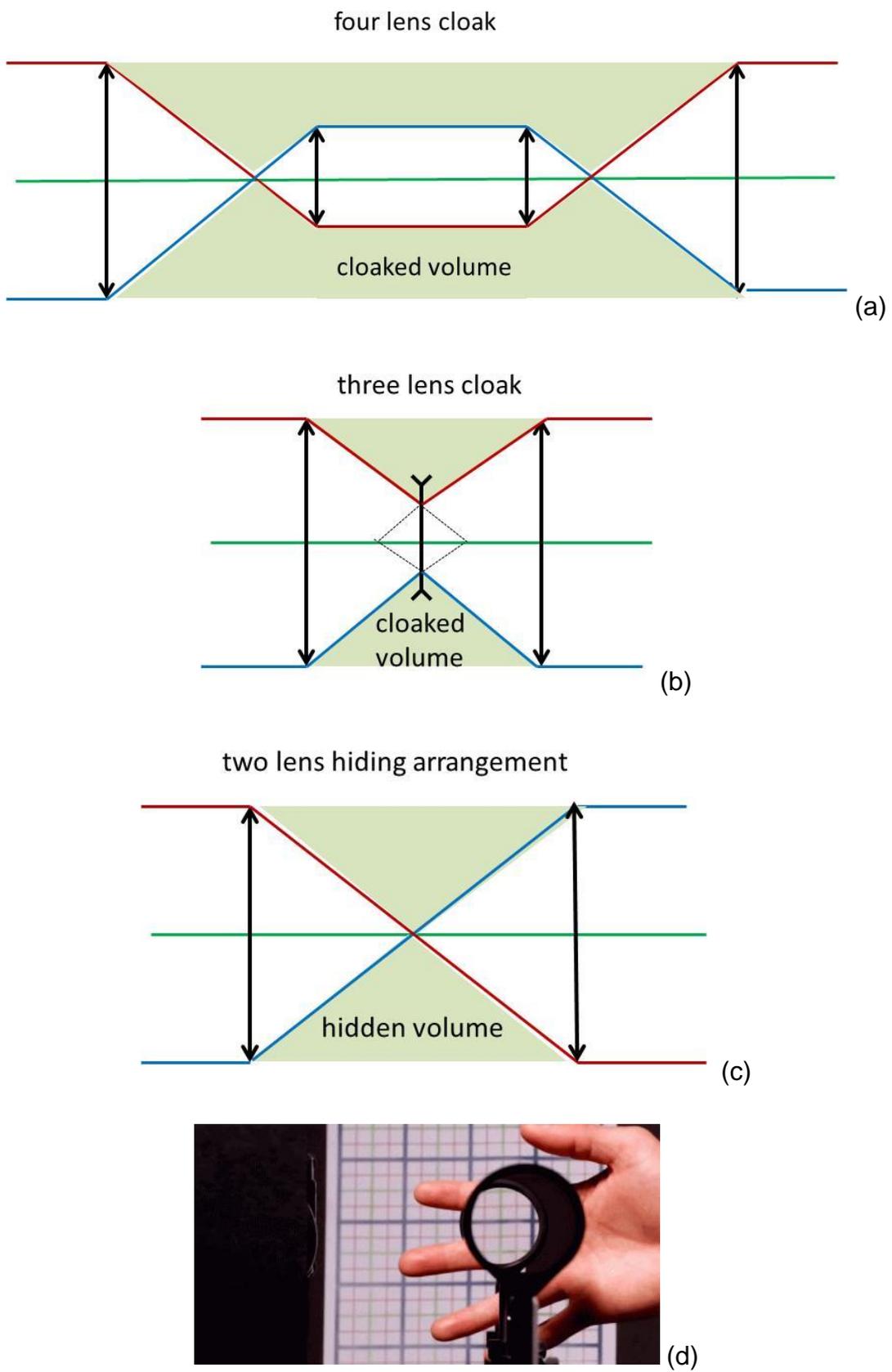

Fig.1

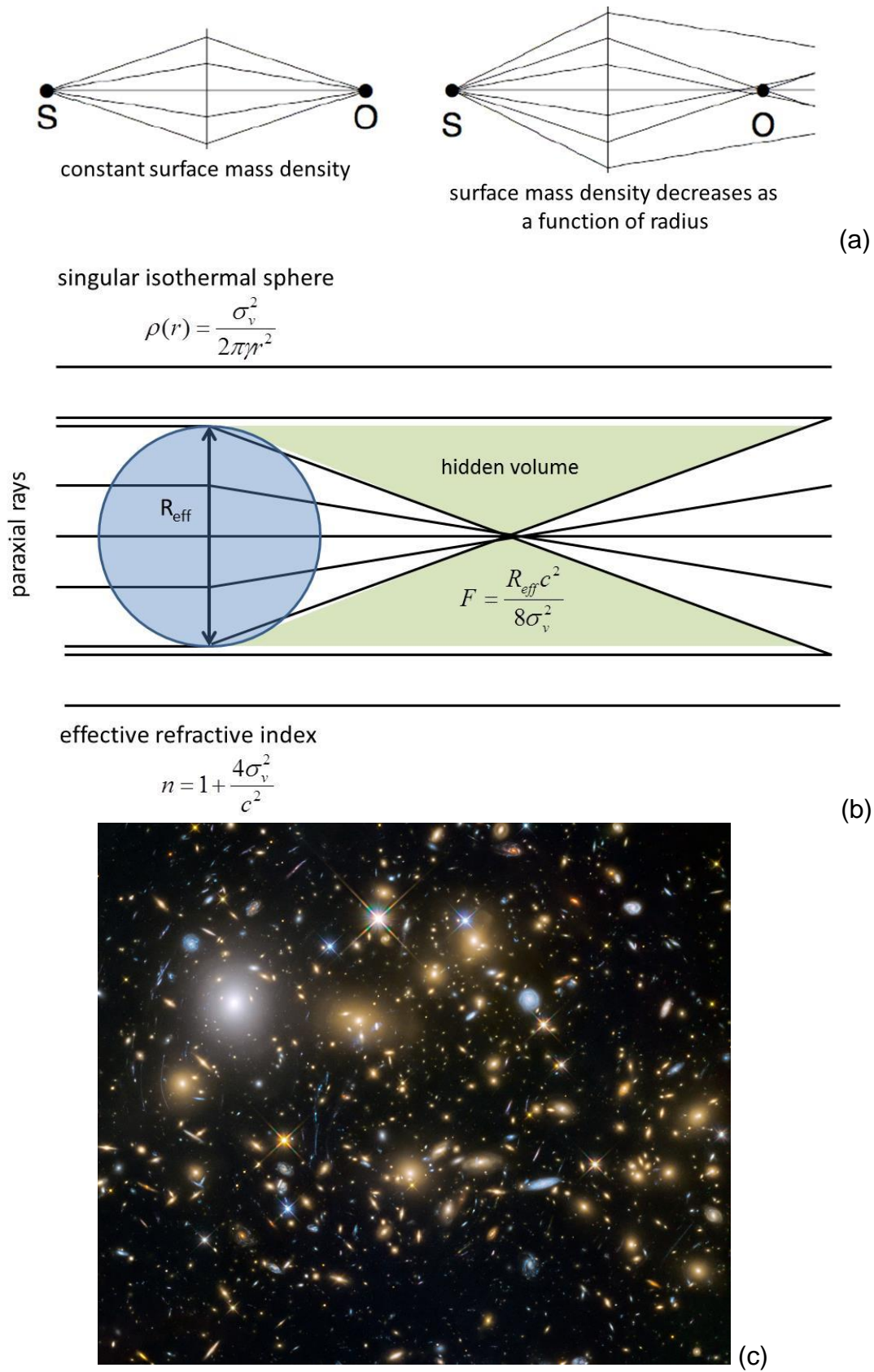

Fig.2

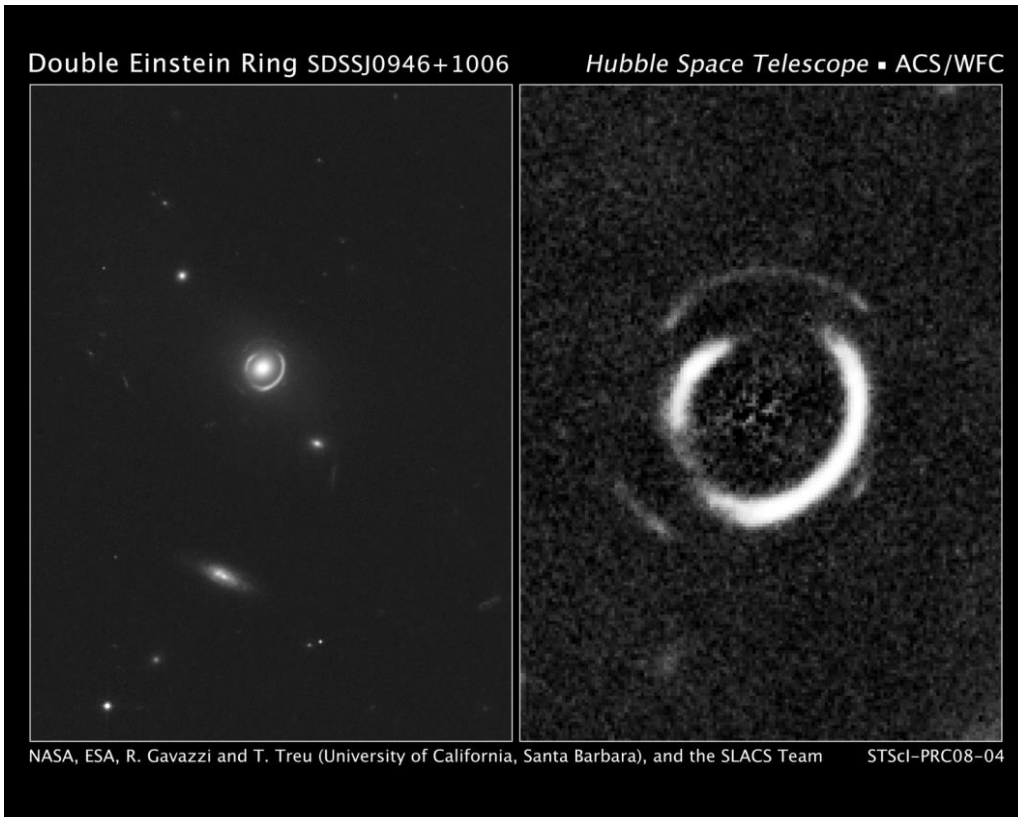

(a)

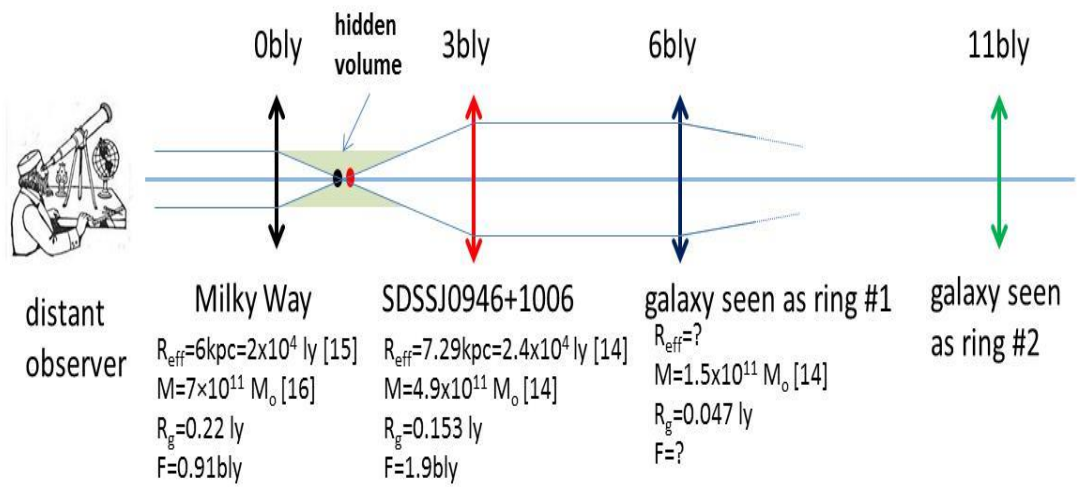

(b)

Fig. 3